\documentclass[aps,prb,twocolumn,superscriptaddress,showpacs]{revtex4}
\usepackage{bm}
\usepackage{amsmath}
\usepackage{amssymb}
\usepackage{graphicx}
\input{epsf}

\begin{document}
\title{Magnetopolaronic Effects in Electron Transport through a Single-Level Vibrating
Quantum Dot}
\author{G. A. Skorobagatko}
\email{gleb_skor@mail.ru}
\affiliation{B. Verkin Institute for Low Temperature Physics and Engineering of
the National Academy of Sciences of Ukraine, 47 Lenin Avenue, Kharkov 61103, Ukraine}
\author{S. I. Kulinich}
\affiliation{B. Verkin Institute for Low Temperature Physics and Engineering of
the National Academy of Sciences of Ukraine, 47 Lenin Avenue, Kharkov 61103, Ukraine}
\affiliation{Department of Physics, University of Gothenburg, SE-412
96 G{\" o}teborg, Sweden}
\author{I. V. Krive}
\affiliation{B. Verkin Institute for Low Temperature Physics and Engineering of
the National Academy of Sciences of Ukraine, 47 Lenin Avenue, Kharkov 61103, Ukraine}
\affiliation{Department of Physics, University of Gothenburg, SE-412
96 G{\" o}teborg, Sweden}
\affiliation{Physics Department, V. N. Karazin National University, Kharkov 61077, Ukraine}
\author{R. I. Shekhter}
\affiliation{Department of Physics, University of Gothenburg, SE-412
96 G{\" o}teborg, Sweden}
\author{M. Jonson}
\affiliation{Department of Physics, University of Gothenburg, SE-412
96 G{\" o}teborg, Sweden}
\affiliation{SUPA, Department of
Physics, Heriot-Watt University, Edinburgh EH14 4AS,
Scotland, UK}
\affiliation{ Division of Quantum Phases
and Devices, School of Physics, Konkuk University, Seoul 143-701, Republic of Korea}

\begin{abstract}
Magneto-polaronic effects are considered in electron transport
through a single-level vibrating quantum dot subjected to a
transverse (to the current flow) magnetic field. It is shown that
the effects are most pronounced in the regime of sequential
electron tunneling, where a polaronic blockade of the current at low
temperatures and an anomalous temperature dependence of
the magnetoconductance are predicted. In contrast, for resonant
tunneling of polarons the peak conductance is not affected by the
magnetic field.
\end{abstract}
 \maketitle

\section{Introduction}

Single-molecule transistors have been intensively studied in recent years
(see, e.g., the review in Ref.~\onlinecite{1}). The specific feature of these
nanostructures is the presence of vibrational effects \cite{2} in
electron transport. Recently, effects of a strong electron-vibron
interaction were observed in electron tunneling through suspended
single wall carbon nanotubes (SWNT) \cite{3,4} and in carbon
nanopeapods.\cite{5} In suspended carbon nanotubes an
electron-vibron coupling is induced by the electrostatic interaction
of the charge on a vibrating molecule with the metal
electrodes. Electrically excited vibrations result in such effects
as phonon-assisted tunneling, \cite{GlazShekh} Franck-Condon
(polaronic) blockade \cite{vonOppen} and electron shuttling
\cite{shuttle} (see also the reviews in Refs.~\onlinecite{2,6,7}).

Much less is known about the influence of a magnetic field on
electron transport in molecular transistors. One can expect that
a magnetic field, interacting with the electric current flowing
through the system, will shift the position of the molecule inside
the gap between the leads. For  point-like electrodes this could
result in a change of electron tunneling probabilities and as a
consequence in a negative magnetoconductance.  For suspended
carbon nanotubes magnetic field-induced displacements (due to
the Laplace force) of the center-of-mass coordinate of the wire does
not influence the absolute values of tunneling matrix elements.
The magnetic influence emerges from a more subtle effect --- the
dependence of the phase  of the electron tunneling amplitude on
magnetic field (Aharonov-Bohm phase). It was shown in
Ref.~\onlinecite{ShGGJ} that despite the 1D nature of electron transport
in SWNTs, a magnetic field applied perpendicular to the quantum wire
results in negative magnetoconductance due to quantum vibrations
of the tube. At low temperatures ($T\ll\hbar\omega_0/k_B$, where $\omega_0$
is the frequency of the vibrational bending mode of the tube) the
conductance, $G(H)$, of the tube is exponentially suppressed,
$G\propto\exp(-\phi^2)$, where $\phi= 2\pi\Phi/\Phi_0$
($\Phi=HLl_0$ is the effective magnetic flux, $L$ is the length of
the wire, $l_0$ is the amplitude of zero-point fluctuations of the
tube, $\Phi_0=hc/e$ is the flux quantum).\cite{ShGGJ} At high
temperatures ($T\gg\hbar\omega_0/k_B$) the magnetic field-induced
correction to the tunnel conductance scales as $1/T$. The scaling properties of
magnetoconductace predicted in Ref.~\onlinecite{ShGGJ} resemble the ones
known for polaronic effects in electron transport through a vibrating quantum qot
(see, e.g., the review in Ref.~\onlinecite{7}) if one identifies the dimensionless flux $\phi_0$
with the electron-vibron coupling constant. So it is interesting
to consider electron transport through a vibrating nano-wire in
the model (single-level QD) when magneto-polaronic effects can be
evaluated analytically. In  Ref.~\onlinecite{Pistol} the resonant
magneto-conductance of a vibrating single-level quantum dot was
calculated in perturbation theory with respect to $\phi\ll 1$ . It was shown in
particular that unlike the case of electrically induced
vibrations, where the peak conductance is known \cite{MAM} to be
unaffected by vibrations, the magnetic field influences resonance
conductance for an asymmetric junction.

The purpose of the present paper is to consider how the electrical current
through a vibrating molecule depends on magnetic field and temperature
in the limit of strong electron-vibron coupling, that is
to consider {\em magneto-polaronic} effects. As in
Ref.~\onlinecite{Pistol} we model the vibrating molecule by a single-level
quantum dot in a harmonic potential. For point-like contacts
both the modulus and the phase of the electron tunneling amplitudes
depend on the QD position in the gap between the leads. We will
assume that the ``longitudinal" center-of-mass coordinate ($x_c$) of the
QD is fixed and that the quantum dot only vibrates along the $y$-direction,
while the magnetic field $H$ is applied along $z$-axis (see Fig.~1).

\begin{figure}
\includegraphics[width=0.95\columnwidth]{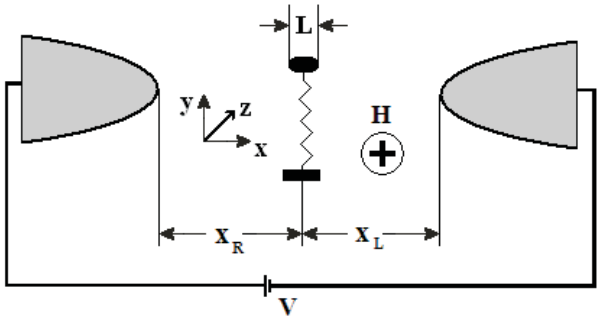}
\caption{Schematic diagram of the device geometry. A single-level
quantum "dot" (QD) of characteristic size $L$ is placed in
the gap between two point-like metal electrodes. The QD position
$x_{L,R}$ on the $x$-axis is fixed and it vibrates (depicted
as a spring) only along the $y$-axis. An external magnetic field is
directed along the $z$-axis.} \label{Fig1}
\end{figure}

At first we calculate the tunneling current in the case when the
displacement $y_c$ of the QD in a magnetic field (due to the
Laplace force \cite{BSL}) is greater than the amplitude of
zero-point fluctuations, $y_c\gg l_0$,
($l_0=\sqrt{\hbar/m\omega_0}$, $m$ is the QD mass). Then the
mechanical part of the problem can be treated  classically and the
dependence of electrical current on magnetic field appears due to
the dependence of the electron tunneling probabilities on the
equilibrium position of the current-carrying QD in the magnetic
field. We show that in strong magnetic fields $H$ (the appropriate
limit for the considered classical problem) the current scales as
$J\propto 1/H$.

In the case when the bare tunneling probabilities do not depend on the
QD displacement $y_c$ ($y_c\ll  x_c$)  the magnetic field
influences the current only through the phase factors
(Aharonov-Bohm phase) in the tunneling Hamiltonian (this is always
the case for a suspended SWNT). We considered magnetopolaronic
effects in the regime of sequential electron tunneling and for
resonant tunneling of polarons (polaron tunneling approximation
\cite{PTA}). In these cases analytical formulae for the current
were derived. We predict: (i) a Franck-Condon (polaronic) blockade
of magneto-conductance in the regime of sequential electron
tunneling, (ii) an anomalous temperature dependence of
magneto-conductance for strong electron-vibron coupling
($\phi\simeq 1$), (iii) a magnetic field-induced narrowing of
resonant conductance peaks, and (iv) an excess current at high biases,
$eV\gg\hbar\omega_0$.

\section{Model Hamiltonian and equations of motion}

The Hamiltonian of a single-level vibrating
 quantum dot in  a magnetic field takes the form
\begin{equation}\label{Ham}
\hat{H}=\sum_{j=L,R}(H_j^{(l)} +  H_j^{(t)}) + H_D ,
\end{equation}
where
\begin{equation}\label{Hl}
H_j^{(l)} = \sum_k(\varepsilon_{k,j}-\mu_j)a_{k,j}^{\dag}a_{k,j}
\end{equation}
is the Hamiltonian of noninteracting electrons in the left ($j=L$)
and right ($j=R$) leads ($\varepsilon_{kj}$ is the energy of
electrons with momentum $k$, $\mu_j$ is the chemical potential),
$a_{k,j}^{\dagger}(a_{k,j})$ are the creation (destruction)
operators  with the standard commutation relations
$\{a_{k,j}^{\dagger},a_{q,j^{\prime}}\}=\delta_{jj^{\prime}}
\delta(k-q)$. Furthermore,
\begin{equation}\label{HD}
H_D=\varepsilon_0c^{\dag}c + \hbar\omega_0b^{\dag}b
\end{equation}
is the Hamiltonian of a single-level ($\varepsilon_0$ is the level
energy) vibrating quantum dot ($\omega_0$ is the frequency of
vibrations in the $y$- direction), $c^{\dag}(c)$ and $b^{\dag}(b)$
are creation (destruction) fermionic ($\{c^{\dag},c\}=1$) and
bosonic ($[b,b^{\dag}]=1$) operators. \cite{2005} Finally,
\begin{eqnarray}\label{Ht}
 H_j^{(t)}& =& t_j(\hat y)\exp(-ij\lambda_H\hat y)a_{k,j}
c^\dag + H.c. \\
 j&=&(L,R)\equiv(-,+)\nonumber
\end{eqnarray}
is the tunneling Hamiltonian. Here $\hat y =
(b^{\dag}+b)\,l_0/\sqrt{2}$ is the coordinate operator
and
$\lambda_H=eHd/\hbar c$ is the magnetic field-induced
electron-vibron coupling \cite{ShGGJ} ($d$ is a parameter of
dimension length;  its physical meaning in our model will be
clarified later). In this section, the amplitude $ t_j(\hat y)$ of the tunneling matrix
element in Eq.~(\ref{Ht}) will be modelled by the expression
\begin{equation}\label{tj}
t_j(\hat y )=t_{0j}\exp\left(-\sqrt{x_j^2+\hat{y}^2}/l_t\right),
\end{equation}
where $l_t$ is the tunneling length and $x_j$ is the position of
the center of mass of QD along $x$-axis, which is assumed to be
fixed. We will show that our results are not sensitive to the
choice of parametrization, Eq.~(\ref{tj}).

The Heisenberg equation of motion for the fermion and boson
operators are
\begin{eqnarray}
&&i\hbar\dot{a}_{kj}=\varepsilon_{kj}a_{kj}+t_j^{\ast}
(\hat{y})\exp(-ij\lambda_H\hat{y})c,\label{eqm1}\\
&&i\hbar\dot{c}=\varepsilon_{0}c+\sum_{k,j}t_j(\hat{y})
\exp(ij\lambda_H\hat{y})a_{kj},\label{eqm2}\\
&&\ddot{\hat{y}}+\omega^2_0\hat{y}= m^{-1}\left(\hat{F}_c +
\hat{F}_L\right),\label{eqm3}
\end{eqnarray}
where the operator expressions for the cohesive ($F_c$) and
Laplace ($F_L$)  forces take the form
\begin{eqnarray}
\hat F_c &=&\sum_{k,j}\frac{\partial t_j(\hat y)} {\partial\hat
y}e^{ij\hat{\phi}}c^{\dagger}a_{kj}+H.c., \label{Fc}\\
\hat{F}_L&=& -\frac{Hd}{c}\left(\hat I_L-\hat
I_R\right). \label{FL}
\end{eqnarray}
In the last equality in Eq.~(\ref{FL}) we introduced the standard
notation for the current operator
\begin{equation}
\hat I_j=\dot N_j,\qquad N_j=\sum_k a_{k,j}^\dagger a_{k,j} .
\end{equation}

At first we neglect the quantum fluctuations of the coordinate
operator $\hat y$ and derive the equation of motion for  the
average (classical) coordinate $<\hat y>=y_c\gg l_{0}$. When QD vibrations
are treated as classical oscillations the equations of motion for
the fermion operators, Eqs.~(\ref{eqm1}) and (\ref{eqm2}), become
a set of first order linear differential equations, which can
easily be solved analytically. After straightforward calculations
(see, e.g., Ref.~\onlinecite{Fed}, where an analogous equation was
derived for the electron shuttle problem) we get the following
classical equation of motion (notice that we made use of the wide
band approximation when calculating the averages over electron
operators and introduced a coordinate-dependent level width
$\Gamma_j(y_c)= 2\pi\nu(\varepsilon_F)|t_j(y_c)|^2$, where
$\nu(\varepsilon_F)$ is the electron density of states in the
leads)
\begin{equation}\label{eqmyc}
\ddot{y_c}+\omega^2_0 y_c= \frac{1}{m}\left(\sum_j
R_j(y_c)\frac{\partial\Gamma_j(y_c)}{\partial y_c}+
\frac{2d}{c}J(y_c)H \right) ,
\end{equation}
where
\begin{equation}\label{Rj}
R_j(y_c) = \frac{1}{2\pi}\int_0^{\infty}d\varepsilon
\frac{(\varepsilon-\varepsilon_0)f_j(\varepsilon)}
{(\varepsilon-\varepsilon_0)^2+[\Gamma_t(y_c)/2]^2}
\end{equation}
and
\begin{equation}\label{Jc}
J(y_c)=\frac{e}{2\pi\hbar}\int_0^{\infty}d\varepsilon
T_{BW}(\varepsilon;y_c)[f_L(\varepsilon)-f_R(\varepsilon)].
\end{equation}
Here $ eV\gg \Gamma_{0} $ $\Gamma_t(y_c)=\Gamma_L(y_c)+\Gamma_R(y_c)$  is the total
level width, and
\begin{eqnarray}\label{TBW}
&&T_{BW}(\varepsilon;y_c)=\frac{\Gamma_L(y_c)\Gamma_R(y_c)}
{(\varepsilon-\varepsilon_0)^2+[\Gamma_t(y_c)/2]^2}\\
&&f_j(\varepsilon)=\left[\exp\left(\frac{\varepsilon-\mu_j}
{T}\right)+1\right]^{-1}
\end{eqnarray}
are the Breit-Wigner transmission cofficient and Fermi-Dirac
distribution function, respectively. Equation (\ref{Jc}) is the standard
Landauer-B\"{u}ttiker formula for the resonant current through a
single-level quantum dot. The first term on the r.h.s of
Eq.~(\ref{eqmyc}) can be interpreted as the cohesive force, the
second term coincides with the force on a current-carrying conductor
in a magnetic field (Laplace force) if we identify $2d$ with the
longitudinal size of the QD, $2d=L$. This is the definition of
the parameter $d$. which appears in the operator form of the Aharonov-Bohm
phase in the tunneling Hamiltonian, Eq.~(\ref{Ht}).

In the absence of a magnetic field, $H=0$, the equilibrium position of
the transverse coordinate $y_c=0$. One can expect that the maximal
influence of the magnetic field on the electrical current through a
single-level QD occurs at high voltages, $eV\geq \Gamma_{L,R}(y_c=0)\equiv
\Gamma_0^{L,R}$, that is  in the regime of sequential electron
tunneling. In this case Eqs.~(\ref{Rj}), (\ref{Jc}) are strongly
simplified and one finds that
\begin{equation}\label{JG}
J(y_c)=\frac{e\Gamma(y_c)}{\hbar},\quad\Gamma(y_c)=
\frac{\Gamma_L(y_c)\Gamma_R(y_c)}{\Gamma_L(y_c)+\Gamma_R(y_c)}
\end{equation}
and ($\varepsilon_F$ is the Fermi energy)
\begin{equation}\label{RLR}
R_L=R_R=\frac{1}{2\pi}\ln\left(\frac{eV}{2\varepsilon_F}\right)
\qquad eV>\Gamma_0^{L,R}.
\end{equation}
For simplification we consider symmetric junctions ($x_L=x_R=l$)
for which
\begin{equation}\label{symj}
\Gamma_L(y_c)=\Gamma_R(y_c)=\Gamma_0\exp\left\{-\frac{2}{l_t}
\left(\sqrt{l^2+y_c^2}-l\right)\right\} ,
\end{equation}
where
$\Gamma_0=2\pi\nu(\varepsilon_F)|t_0|^2\exp(-2l/l_t)$ is the
level width of the symmetric junction in the absence of a magnetic field.
According to Eq.~(\ref{eqmyc}) the equilibrium position of the QD in a
constant magnetic field does not depend on time; in weak magnetic
fields it scales linerly with $H$,
\begin{equation}\label{yw}
y_c(H)\simeq \frac{L}{2}\frac{H}{H_0}\ll l,\quad
H\ll\frac{l}{L}H_0 ,
\end{equation}
where the characteristic magnetic field $H_0$ is defined by the
equation
\begin{eqnarray}
&&\frac{I_0H_0}{c}=m\omega_R^2,\quad I_0=\frac{e\Gamma_0}
{\hbar},\label{H0}\\
&&\omega_R^2=\omega_0^2+\frac{2}{\pi}\frac{\Gamma_0}{mll_t}\ln
\left(\frac{2\varepsilon_F}{eV}\right).\nonumber
\end{eqnarray}

We see from Eqs.~(\ref{yw}) and (\ref{H0}) that in weak magnetic fields
the only effect of the cohesive force is to renormalize the frequency
$\omega_0$ . In tunnel junctions ($\Gamma_0\rightarrow 0$) the
renormalization is small and can be neglected. In strong magnetic
fields, $H\gg(l/L)H_0$, we can neglect the contribution of the cohesive
force to Eq.~(\ref{eqmyc}) as well. The magnetic field-induced
shift of the QD in this limit scales logarithmically with $H$,
\begin{equation}\label{yst}
y_c(H)\simeq \frac{l_t}{2}\ln\left\{\frac{L}{2l}
\frac{H}{H_0}\exp\left(\frac{2l}{l_t}\right)\right\} \gg l  .
\end{equation}

In weak magnetic fields the small shift ($y_c\ll l$) of the QD position 
does not influence the tunnel current in the
considered classical approach. We will see in the next section
that in this case one has to take into account quantum effects
(phase fluctuations in the tunnelling Hamiltonian), which strongly
modify tunnel transport. In strong magnetic fields (classical
limit) the current $J(H)=(e/\hbar)\Gamma(H)$ scales
as $1/H$ according to Eq.~(\ref{yst})  .

At the end of this Section we briefly comment on the influence of
the magnetic field on the resonant current in the considered classical
approach. In the regime of resonant electron tunneling ($T, eV\leq
\Gamma_0$)  the current depends linearly on the bias voltage $V$,
\begin{equation} \label{Jr}
J_r(H) = 4G_0\frac{\Gamma_L(H)\Gamma_R(H)}{[\Gamma_L(H)+\Gamma_R(H)]^2} V
\end{equation}
($G_0=e^2/h$ is the conductance quantum). It is evident from
Eq.~(\ref{Jr}) that the  resonant current through a symmetric
junction ($\Gamma_L=\Gamma_R$) is not affected by the magnetic field since
$J_r(H)=J_r(0)=G_0V$. For an asymmetric junction the resonant current
does not depend on magnetic field  in the strong-$H$ limit when
the field-induced factor in the expression for the renormalized partial widths (see
Eq.~(\ref{symj})) is cancelled in the expression for the electrical
current, Eq.~(\ref{Jr}).

\section{Magneto-polaronic effects.}

In this section we consider the influence of quantum and
thermodynamical fluctuations of the coordinate operator of QD,
$\hat{y}$, on electron transport in a magnetic field. Quantum effects
are significant (at low temperatures) when one can neglect the
dependence of the modulus of the tunneling matrix element on
the QD displacement in a magnetic field. This is always the case for
tunneling through a suspended SWNT.\cite{ShGGJ}

When considering the quantum effects of magnetic-field induced
vibrations it is convenient to introduce the electron-vibron
coupling constant in the form of a dimensionless magnetic flux,
$\phi=2\pi\Phi/\Phi_0$ ($\Phi=HLl_0/2$, 
$\Phi_0=hc/e$ is the flux
quantum). The dimensionless electron-vibron interaction constant
$\phi$ determines the quantum phase of the tunneling matrix element
\begin{equation} \label{ABph}
t_{L/R}(\hat{y}) = t_{0j}\exp[\mp i\phi(b^{\dagger}+b)/\sqrt{2}],
\end{equation}
where $l_0=\sqrt{\hbar/m\omega_0}$ is the amplitude of zero-point
fluctuations. Resonant electron tunneling in the model
Eqs.~(\ref{Ham})-(\ref{Ht}), (\ref{ABph}) was studied in
Ref.~\onlinecite{Pistol} using perturbation theory with respect to $\phi\ll 1$. Here we
are interested in non-perturbative effects, $\phi\simeq 1$.

At first we consider the regime of sequential electron tunneling
where the effects of magnetic field-induced vibrations are most
pronounced. In this regime the current can be calculated
perturbatively with respect to the level width $\Gamma$. The sign of
the Aharonov-Bohm phase, Eq.~(\ref{ABph}), which is opposite for left-
and right-tunneling electrons, does not play any role in the
considered regime of tunneling (which can be treated classically
by using a master equation approach). So our model is equivalent to
the polaronic model of electron tunneling through a vibrating QD
(see, e.g., the reviews in Refs.~\onlinecite{2,7}). Notice that in a general case the
``magnetic"  problem can not be mapped to the polaronic problem
because of the  above mentioned ``sign" difference. \cite{Pistol} We
show below that this difference is not essential for
magneto-polaronic effects.

The average electric current $J$ in the regime of sequential
electron tunneling ($k_BT, eV\gg\Gamma$) can be calculated by using
a master equation approach. It can be represented as a sum of
partial currents over ``vibron channels" \cite{LM}, $n>0 (n<0)$,
corresponding to vibron emission (absorption). Hence,
\begin{equation} \label{Jseq}
J(V;\phi)=J_0\sum_{n=-\infty}^{\infty}
A_n(\phi)\{f_L(\varepsilon_0-n\hbar\omega_{0})-f_R(\varepsilon_0-n\hbar\omega_{0})\},
\end{equation}
where $J_0=e\Gamma/\hbar$ is the maximal current through a
single-level QD ($\Gamma=\Gamma_L\Gamma_R/(\Gamma_L+\Gamma_R$))
and the spectral weights $A_n$ are defined by the equation
\begin{equation} \label{sw}
\sum_{n=-\infty}^{\infty}A_ne^{in\omega_0t} =
\left<e^{\pm i\phi\hat{y}(t)}e^{\mp i\phi\hat{y}(0)}\right> ,
\end{equation}
where the average $<...>$ is taken with respect to the Hamiltonian
of noninteracting vibrons, $H_b=\hbar\omega_0b^{\dag}b$. The
spectral weights defined by Eq.(\ref{sw}) coincide with the
analogous quantities in the polaronic model, where they are
defined through the correlation function of operators
$\exp(i\lambda \hat{p})$ ($\hat{p}$ is the momentum operator).
For the equilibrated vibrons with the distribution function
$n_B(T)=[\exp(\hbar\omega_0/T)-1]^{-1}$ the coefficients  $ A_n$
take the following well-known form (see, e.g., Ref.~\onlinecite{Mahan})
\begin{eqnarray} \label{sw1}
&&A_n(\frac{\hbar\omega_0}{T};\phi) = \exp[-\phi^2(1+2n_B)]
\times\nonumber\\
&&I_n[2\phi^2\sqrt{n_B(1+n_B)}]\exp
\left(-n\frac{\hbar\omega_0}{T}\right).
\end{eqnarray}
Here $I_n(z)$ denotes a modified Bessel function (see, e.g.,
Ref.~\onlinecite{RG}). As is evident from  their definition in Eq.~(\ref{sw}),
the spectral weights $A_n$ satisfy the sum rule
\begin{equation} \label{sumr}
\sum_{n=-\infty}^{\infty}A_n = 1 ,
\end{equation}
which can be rewritten as a nontrivial mathematical identity for
the sum of modified Bessel functions,
\begin{equation} \label{sBf}
\sum_{n=-\infty}^\infty e^{-nx}I_n\left(\frac{a}{\sinh x}\right)
=e^{-a\coth x},\quad a\geq 0.
\end{equation}
The unitary condition Eq.(\ref{sumr}) plays a crucial role in the
derivation of analytical formulae for the current and conductance
at $k_BT, eV\gg\hbar\omega_0$. Since all the
analytical formulas of interest for us have already been derived
in the literature on the polaronic model, we here merely formulate
the results.

The magneto-conductance $G(H)$ in the regime of sequential
electron tunneling takes the following asymptotics at low and high
temperatures  \cite{KF}
\begin{equation} \label{amc}
\frac{G(H)}{G(0)}\simeq\left\{
\begin{array}{ll}
\exp(-\phi^2), & \Gamma\ll T\ll\hbar\omega_0 \\
1-\phi^2\frac{\hbar\omega_0}{2T} , & T\gg \phi^2\hbar\omega_0
\end{array}\right.
\end{equation}
Here $G(0)=G_0(\pi/2)(\Gamma/T)$ is the standard formula for
conductance of a single-level QD. At intermediate temperatures,
$k_BT\sim\hbar\omega_0$ and for strong electron-vibron interactions,
$\phi\simeq 1$, the temperature dependence of conductance is
nonmonotonic (anomalous).\cite{KF} This signature of polaronic
effects was observed in experiments\cite{5} on electron tunneling
through a carbon nanopeapod-based single-electron transistor. The
asymptotics Eq.~(\ref{amc}) coincide (up to numerical factors) with
the ones found in Ref.~\onlinecite{ShGGJ} for a different model (electron
tunneling through a suspended nanotube). The high temperature
asymptotics in  Eq.~(\ref{amc}) exactly coincides with the
corresponding quantity calculated in Ref.~\onlinecite{Pistol} for
resonant electron tunneling. It is interesting to notice that the
calculations based on a full quantum mechanical treatment
\cite{Pistol} of interacting electrons and the master equation
approach yield exactly the same results in high-$T$ limit.

Now we consider the behavior of current, Eq.~(\ref{Jseq}), at low
temperatures as a function of bias voltage ($eV\gg\Gamma$). At
$T=0$ Eq.~(\ref{Jseq}) takes the form
\begin{equation} \label{JT0}
J(V;\phi)=J_0\exp(-\phi^2)\sum_{n=0}^{n_m}\frac{\phi^{2n}}{n!} ,
\end{equation}
where $n_m=[eV/\hbar\omega_0]$ ( $[x]$ denotes the integer part of
$x$).  At low voltages ($eV<\hbar\omega_0$ ) $n_m=0$ and the
current does not depend on $V$ and is determined by a standard
formula for a saturated current through a single level QD.
However, in our case the level width $\Gamma(\phi)$ is
renormalized by the electron-vibron interaction,
\begin{equation} \label{Gphi}
J(V<\hbar\omega_0/e;\phi)\simeq\frac{e\Gamma(\phi)}{\hbar}=
\frac{e\Gamma}{\hbar}\exp(-\phi^2) .
\end{equation}
This is the demonstration of polaronic blockade. \cite{vonOppen}
With an increase of bias voltage the current jumps by an amount
determined by the Franck-Condon factors
\begin{equation} \label{FCf}
\Delta J_n=J_{n+1}-J_n=J_0e^{-\phi^2}\frac{\phi^{2n}}{n!}
\end{equation}
each time the bias voltage opens a new inelastic channel
$n(V)=[eV/\hbar\omega_0]$, (see Fig.~2).
\begin{figure}
\includegraphics[width=1.0\columnwidth]{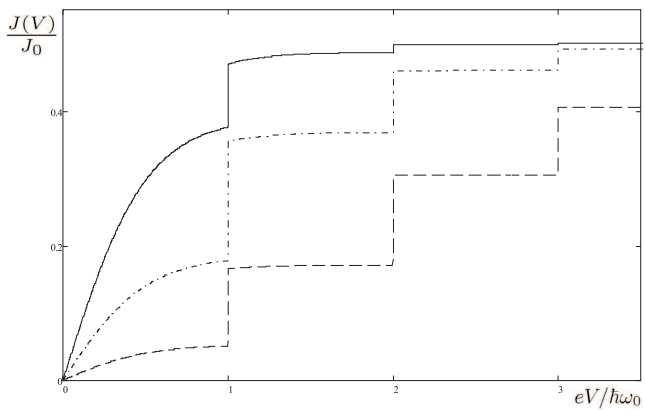}
\caption{Low-temperature current $J(V)$ in units of
$J_0=e\Gamma/\hbar$ as a function of bias voltage at: $\beta=\hbar\omega_0/T=4$, for three
different values of the electron-vibron coupling constant $\phi$: the
solid line corresponds to the weak- ($\phi=0.5$), the dash-dotted line
to an intermediate- ($\phi=1$), and the dashed line to the strong
($\phi=1.5$) coupling regimes.} \label{Fig2}
\end{figure}

At high voltages ($eV\gg\hbar\omega_0$) the polaronic blockade is
lifted and the current saturates at its maximum value $J_0$,
\begin{equation} \label{satur}
J(V\gg\hbar\omega_0/e;\phi)\simeq J_0\left\{1-e^{-\phi^2}
\frac{\phi^{2n_m}}{\Gamma(n_m+1)}\right\} .
\end{equation}
Here $\Gamma(x)$ is the gamma-function and $n_m=[eV/\hbar
\omega_0]\gg\phi^2\gg 1$. The difference between the maximum and
minimum currents
\begin{equation} \label{excess}
\delta J = J_0\{1-\exp(-\phi^2)\}
\end{equation}
is nothing but the excess current considered in
Refs.~\onlinecite{Sonne1,Sonne2} for the model of electron tunneling
through a suspended carbon nanotube. The  presence of a high-temperature
($k_BT\ll eV\rightarrow\infty$) excess current in
electron tunneling is another demonstration of polaronic blockade
effects.

Although the magnetic field-induced polaronic effects are most
pronounced in the regime of sequential electron tunneling (when
the current through a single-level QD is maximal) we briefly
comment here on polaronic effects in resonant electron tunneling.
It is physically evident that an electron-vibron polaron state can be
formed on a quantum dot coupled to reservoirs if the life-time of
the electronic state, $\tau\sim \hbar/\Gamma_t$, is much longer
than the characteristic time of polaron formation,
$\tau_p\sim\hbar/\varepsilon_p\sim 1/\omega_0\phi^2, (\phi\geq
1)$. The corresponding inequality $\Gamma_t\ll\phi^2\hbar\omega_0$ allows one
to consider resonant tunneling of strongly interacting electrons
in a simple model (polaron tunneling approximation \cite{PTA}). In
this approximation the bare electron Green's function (GF) in the
Dyson equation for  the retarded (advanced) GFs  is replaced by
the polaron GF ($G_p(\varepsilon)$)
\begin{equation} \label{Dyson}
G_{r,a}^{(PTA)}=\left[G_p^{-1}(\varepsilon) -
\Sigma_{r,a}\right]^{-1},
\end{equation}
where
\begin{equation} \label{Gp}
G_p(\varepsilon) = \sum_{n=-\infty}^\infty\frac{A_n}
{\varepsilon-\varepsilon_0+n\hbar\omega_0}
\end{equation}
and $A_n$ are defined in Eq.~(\ref{sw}).  In the limit of wide
electron bands in the leads the imaginary part of the self-energy
function $\Sigma(\varepsilon)$ is not renormalized by
electron-vibron interaction in the considered approach
$\text{Im}\Sigma_{r,a}=\mp\Gamma_t/2$ and the real part of
$\Sigma(\varepsilon)$ can be neglected. Then by evaluating the
spectral function $A^{(PTA)}=i[G_r^{(PTA)}-G_a^{(PTA)}]$ one can
find the current with the help of the Meir-Wingreen formula \cite{MW}.
It takes the form \cite{PTA}
\begin{equation} \label{Tp}
J=\frac{e}{\hbar}\int d\varepsilon\frac{\Gamma_L\Gamma_RG_p^2
(\varepsilon)}{1+[\Gamma_tG_p(\varepsilon)/2]^2}
\left\{f_L(\varepsilon)-f_R(\varepsilon)\right\} .
\end{equation}
From Eqs.~(\ref{Gp}),(\ref{Tp}) it is easy to show that at low
temperatures ($T\ll\hbar\omega_0$) the conductance
$G(T,\varepsilon_0)$ in resonant tunneling can be represented in
the Breit-Wigner form with the renormalized level widths
$\Gamma_{L,R}(\phi)=\exp(-\phi^2)\Gamma_{L,R}
(\Gamma_{L,R}\ll\phi^2\hbar\omega_0)$:
\begin{equation} \label{Gres}
G(0)=G_0\frac{\Gamma_L(\phi)\Gamma_R(\phi)}
{(\varepsilon_F-\varepsilon_0)^2+[\Gamma_L(\phi)+\Gamma_R(\phi)]^2/4}.
\end{equation}
According to Eq.~(\ref{Gres}) the peak conductance, $G_r(0,
\varepsilon_0=\varepsilon_F)$, is not
renormalized by the magnetic field even for an asymmetric junction,
$G_r=4G_0\Gamma_L\Gamma_R/[\Gamma_L+\Gamma_R]^2$, as is the case in the
polaronic model. \cite{MAM} Notice that the opposite statement,
that in an asymmetric junction the peak conductance is influenced by
a magnetic field \cite{Pistol}, was obtained in perturbation theory
with respect to the electron-vibron coupling constant $\phi\ll 1$ and that it holds in
another region of model parameters, $
\Gamma_{L,R}\gg\phi^2\hbar\omega_0$.

At high temperatures ($k_BT\gg\phi^2\hbar\omega_0$) the polaronic
blockade is lifted and the formula for the conductance derived
from Eq.~(\ref{Tp}) coincides with the corresponding formula (see
Eq.~(\ref{amc})) obtained in the regime of sequential electron
tunneling.

\section{Conclusion.}

In conclusion we have shown that the quantum-vibration-induced
Aharonov-Bohm effect, predicted in Ref.~\onlinecite{ShGGJ} for
electron tunneling through a suspended carbon nanotube in magnetic
field, can be interpreted as a magneto-polaronic effect, where the
dimensless flux $\phi=2\pi\Phi/\Phi_0$ plays the role of a
magnetic field-induced electron-vibron interaction constant. We
considered a simple model in the form of a single-level vibrating
quantum ``dot" (QD) in a transverse (with respect to the current
flow)  magnetic field - and evaluated the electrical current and
the magnetoconductance in two cases: (i) the amplitude of electron
tunneling depends on the magnetic field-induced QD displacement
(point-like contacts), and  (ii) the magnetic field influences
only the Aharonov-Bohm phase of the tunneling matrix element. It
was shown that magnetic field-induced polaronic effects are most
pronounced: (i) in the regime of sequential electron tunneling,
(ii) in high magnetic fields when the momentum $\delta p\sim
eHL/c$ of the current-carrying QD induced by the Laplace force
exceeds the momentum of zero-point fluctuations
$p_0\sim\hbar/l_0$, and (iii) at low temperatures, $k_BT\ll\delta
p^2/m$ ($m$ is the mass of the QD).

Recently polaronic effects were measured in nanotube-based
single-electron transistors.\cite{3,4,5} In particular, a Franck-Condon
blockade was observed in a suspended carbon nanotube.\cite{4}
Electrically induced electron-vibron interactions happen to be
much stronger than the electron-phonon interaction in isolated carbon
nanotubes. So the magnetic effects could also be enhanced in the
presence of ferromagnetic leads. Although simple estimations for
a micron-sized nanotube-based device show  that even in a very
strong transverse magnetic field ($H\sim 20 T$) the magnetocurrent
is only of the order of 0.1 pA, the effect is measurable and its
fundamental nature justifies efforts to detect it.

\section{Acknowledgement.}

The authors thank  L. Y. Gorelik and F. Pistolesi for valuable
discussions. Financial support from the European Commission
(FP7-ICT-FET Proj. No. 225955 STELE), the Swedish VR, the
Korean WCU program funded by MEST/NFR (R31-2008-000-10057-0) and
the  Grant ``Quantum phenomena in nanosystems and nanomaterials at
low temperatures" (No. 4/10-H) from the National Academy of Sciences
of Ukraine is gratefully acknowledged. I.V.K. and S.I.K.
acknowledge the hospitality of the Department of Physics at the
University of Gothenburg.


\begin{thebibliography}{99}
\bibitem{BSL}
Each of the moving charges that contribute to the current through
a conductor experiences a Lorentz force in an electromagnetic
field. The Laplace force refers to the total force on the
conductor.
\end{thebibliography}
\end{document}